# Gated Tuned Superconductivity and Phonon Softening in Mono- and Bilayer MoS$_2$

Yajun Fu[1]*, Erfu Liu[1]*, Hongtao Yuan[2,3]*, Peizhe Tang[3]*, Biao Lian[3], Gang Xu[3], Junwen Zeng[1], Zhuoyu Chen[3], Yaojia Wang[1], Wei Zhou[1], Kang Xu[1], Anyuan Gao[1], Chen Pan[1], Miao Wang[1], Baigeng Wang[1], Shou-Cheng Zhang[2,3], Yi Cui[2,3], Harold Y. Hwang[2,3] & Feng Miao[1]

[1]National Laboratory of Solid State Microstructures, School of Physics, Collaborative Innovation Center of Advanced Microstructures, Nanjing University, Nanjing 210093, China.

[2]Stanford Institute for Materials and Energy Sciences, SLAC National Accelerator Laboratory, Menlo Park, California 94025, USA.

[3]Geballe Laboratory for Advanced Materials, Stanford University, Stanford, California 94305, USA.

* These authors contributed equally.

**E-mail:** miao@nju.edu.cn (F. M.); bgwang@nju.edu.cn (B. W.); htyuan@stanford.edu (H. T. Y.)

**Abstract**

Superconductors at the atomic two-dimensional (2D) limit are the focus of an enduring fascination in the condensed matter community. This is because, with reduced dimensions, the effects of disorders, fluctuations, and correlations in superconductors become particularly prominent at the atomic 2D limit; thus such superconductors provide opportunities to tackle tough theoretical and experimental challenges. Here, based on the observation of ultrathin 2D superconductivity in mono- and bilayer molybdenum disulfide ($MoS_2$) with electric-double-layer (EDL) gating, we found that the critical sheet carrier density required to achieve superconductivity in a monolayer $MoS_2$ flake can be as low as $0.55 \times 10^{14}$ cm$^{-2}$, which is much lower than those values in the bilayer and thicker cases in previous report and also our own observations. Further comparison of the phonon dispersion obtained by *ab initio* calculations indicated that the phonon softening of the acoustic modes around the M point plays a key role in the gate-induced superconductivity within the Bardeen–Cooper–Schrieffer (BCS) theory framework. This result might help enrich the understanding of 2D superconductivity with EDL gating.

**Keywords:** 2D superconductivity, $MoS_2$, electric-double-layer gating, Phonon softening

**Introduction**

Discovery of the interfacial superconducting state at the atomic scale has motivated the pursuit of emergent condensed phases in two-dimensional (2D) electronic systems. Studies of interfacial superconductivity have been generally limited to the regime in which the superconducting order parameter is restricted to 2D. Examples include one-atomic-layer Pb or In films on a Si (111) substrate,[1-5] single-unit-cell FeSe films on a $SrTiO_3$ substrate [6-9] and few layer $NbSe_2$ crystal.[10, 11] Recent advances in electric-double-layer (EDL) gating have enabled the continuous tuning of one of the order parameters—the superfluid density—in 2D superconductors on the surface of bulk crystals with unprecedented control of surface band bending, doping level and vortex interaction, thus opening up new opportunities for understanding 2D superconductivity in the surface accumulation layers.[12-18] With the reduced dimensionality—especially in a strict 2D monolayer form in layered materials in which the disorder, fluctuation and correlation effects all play particularly important roles—being the key parameter, how low the carrier density can be to realize the interfacial superconductivity approaching the ultimate atomic limit remains elusive.

Therefore, to investigate the superconductivity at the monolayer limit starts to be a topic of great interest.[18] As a representative semiconducting layered material, mono- and bilayer $MoS_2$ (Figure 1a) was chosen here as an ideal platform to study superconductivity at its 2D limit because of the following advantages: 1. The dramatic change in the band structure once the crystal is thinned down from bilayer to monolayer provides us with the possibility to understand the effect of some specific subband structures and Fermi surfaces on the superconductivity; 2. The unique electric-field-driven Zeeman splitting [16, 19, 20] in the valleys of conduction bands might potentially provide spin-valley locking or a triplet electron pairing mechanism in the unique band structure of monolayer $MoS_2$; [21] and 3. Technically, relatively clean atomically thin mono- and bilayer flakes are easily accessible by using mechanical cleavage. In addition, combined with the powerful accessibility of the EDL gating technique, superconductivity in mono- and bilayer $MoS_2$ can help us achieve a deep understanding of the microscopic mechanism of 2D superconductivity. Recently, Costanzo *et al* [18] used this technique to study the superconductivity in thin $MoS_2$ flakes with thickness ranging from mono- to six-layers.

In this work, we also demonstrated gate-induced 2D superconductivity in mono- and bilayer $MoS_2$ in an electric-double-layer transistor (EDLT) geometry and we found that the monolayer sample showed a smaller critical sheet carrier density requirement to achieve superconductivity than those of bilayer and bulk samples. The comparison studies of their vortex motions on monolayer and bilayer superconductivity were conducted to gain insight into the differences in the microscopic pictures of the 2D superconductivity. Our further *ab initio* calculations identified that the superconductivity are mainly induced by the phonon-softening of in-plane acoustic phonon modes and via such a mechanism the monolayer $MoS_2$ can be more easily driven into superconducting phase with the less electron doping compared to the bilayer case. This work might enrich the understanding of gated interfacial superconductivity approaching the ultimate atomic limit and provides a method to achieve new types of superconductors.

**Results**

Figures 1b and 1c show the typical device structure (optical image) and measurement geometry (schematic illustration) of EDLT using an ionic liquid (DEME-TFSI) as the gate medium. Due to the large tunability of the chemical potential using EDLTs, Figure 1d presents an ambipolar operation of bilayer MoS$_2$ EDLT devices at 220 K with an ON-OFF ratio of $10^5$ and a maximum attainable sheet carrier density ($n$) up to ~$2 \times 10^{14}$ cm$^{-2}$ by applying sufficiently high gate voltage $V_{LG}$ via the EDL gating medium. As shown in Figure 1e, in a typical case of a bilayer MoS$_2$ device with $n = 1.3 \times 10^{14}$ cm$^{-2}$ (measured at 10 K), clear superconducting behavior can be observed. The superconducting state was gradually suppressed upon the application of an increasing perpendicular magnetic field and was completely suppressed when the magnetic field reached 1 T (see supplementary information for more details). In addition to the main drop in sheet resistance $R_s$ at 3.3 K, we noted two other small resistance drops at higher temperatures above the superconducting transition, which are likely caused by the slightly inhomogeneous carrier accumulation. These observations suggest that the inhomogeneity-induced fluctuation in the chemical potential of the channel could cause a more notable effect on the electron transport properties while approaching the 2D limit.

To examine the dimensionality nature of such EDL gating-induced superconductivity, we performed 4-probe voltage-current (*V-I*) measurements and found that the results satisfied the Berezinskii-Kosterlitz-Thouless (BKT) transition [22-24] (supplementary Figures S3, S4) for 2D superconductivity. The BKT transition corresponds to the spontaneous dissociation of vortex-antivortex pairs into free vortices at the transition temperature $T_{BKT}$. By fitting the $V$-$I$ and $R_s$-$T$ curves, we obtained a transition temperature $T_{BKT}$ ~2 K (see supplementary Figures S4 for more details). Compared with thicker $MoS_2$ devices (~20 nm),[14] enhancement of back gate modulation due to the atomically thin geometry was also observed (supplementary Figure S5).

Superconductivity in a monolayer $MoS_2$ flake is shown in Figure 1f. The magnetic field response (shown in Figure 1f) confirmed that the resistance drop during cooling-down is most likely the superconducting transition. Two remarkable characteristic features must be addressed here. First, before the transition to a superconductor, the resistance of both mono- and bilayer flakes increases with decreasing temperature (more pronounced in the monolayer case), showing a stronger insulating behavior compared with that observed in thicker $MoS_2$ flakes (20 nm).[14] This result indicates that disorder plays a more important role in the 2D phase transition while approaching the thin limit. This disorder might originate from the inhomogeneous charge accumulation or the substrate effect on the channels. Second, the critical carrier density required to achieve the superconducting state in the monolayer case is much lower than those values observed in our bilayer case or reported thicker cases,[14, 16, 20] which we will discuss in detail below. We note that due to the relatively small applied gate voltage to avoid possible electrochemical reactions and the limited temperature accessibility of our equipment, a zero resistance superconducting state was not reached. However, the interface effect between electrodes and sample, [25, 26] finite-size effects [24] and slightly inhomogeneous superconductivity [15, 27] may also play roles in the non-zero resistance of low dimensional superconductivity. The bilayer superconductivities with zero resistance state under similar conditions suggest that the non-zero resistance case in monolayer $MoS_2$ cannot be explained by finite-size effects or interface effect between electrodes and sample. The small applied gate voltage and slightly inhomogeneous superconductivity may be possible reasons and will require further investigations.

To study the detailed differences in the critical carrier densities required to achieve superconductivity between mono- and bilayer $MoS_2$, we first measured the low-temperature transport properties of bilayer devices at different carrier densities. Figure 2a shows the $R_s$-$T$ curves ($T$ = 200 K to 1.6 K) of a different bilayer device (which is different to the one shown in Figure 1e). With $n$ = 0.95 × $10^{14}$ cm$^{-2}$ (measured at 10 K with Hall effect results shown in the inset of Figure 2a), the device remained metallic until a metal-insulator transition appeared at temperatures down to ~11 K, and no signature of superconductivity was observed. For the same devices, once the sheet carrier density $n$ was increased slightly to ~1.23 × $10^{14}$ cm$^{-2}$, a sharp $R_s$ drop emerged at temperatures of ~4.2 K, and the device reached a zero resistance state at ~1.55 K, indicating the occurrence of superconductivity. For a monolayer $MoS_2$ device, however, a much smaller critical carrier density (~0.55 × $10^{14}$ cm$^{-2}$, with the results shown in Figure 2b) was observed than for those observed in bilayer devices. We summarize the main results in Figure 2c, where $T_c$ [defined as $R_s(T_c)$ = 0.9$R_s$(10

K)] is plotted as a function of $n$ for several typical devices (see a similar plot using number of carrier per primitive cell as the standard of the carrier density in Figure S6, and additional data on homogeneity of carrier densities for multiple pairs of contacts in Figure S7). All the carrier densities were determined by measuring Hall effect at 10 K, where the mono- and bilayer $MoS_2$ maintained normal states. The Hall resistance, $R_{xy}$, show antisymmetric and linear characteristics when plot as a function of magnetic field (such as the inset of Figure 2a and 2b). The negative sign of $R_{xy}$ for positive magnetic field indicate electron-type carriers, which is consistent with the positive gate biases. And the carrier densities were extracted from the Hall coefficient $R_H$ for each gate voltage by using the formula $R_H = 1/ne$. Because disorders or fluctuations are generally believed to play a crucial role in disturbing the superconductivity upon approaching the 2D limit, it is counterintuitive to find that in the ultimate limit case (monolayer), it is "easier" to realize superconductivity than in the bilayer or bulk cases when tuning such a key parameter—carrier density $n$.

Another approach to gain insight into the differences in the microscopic pictures of the 2D superconductivity in mono- and bilayer $MoS_2$ devices is to study their vortex motion.[17] Through analysis of $R_s$-$T$ plots under various perpendicular magnetic fields on typical bilayer (Figure 3a) and monolayer (Figure 3b) devices, we observed the activated behavior of the vortex dynamics at temperatures slightly lower than $T_c$ and at the resistance saturation at even lower temperatures. This behavior is similar to that recently observed in a ZrNCl EDLT superconductor[17] and in disordered metal thin films.[28] At temperatures slightly lower than $T_c$, the sheet resistance can be described

by

$$R_S = R_0(H)e^{-\frac{U(H)}{k_B T}}, \tag{1}$$

where $U(H)$ is the activation energy, and $k_B$ is Boltzmann's constant. The fitting (black dotted lines in Figures 3a and 3b) yields values of $U(H)$. We further plotted the dependence of $U(H)/k_B$ on $H$, as shown in Figure 3c, and found that both mono- and bilayer devices follow the relation of

$$U(H)=U_0\ln(H_0/H), \tag{2}$$

where $U_0 \sim \Phi_0^2 d/256\pi^3\lambda^2$ represents the vortex-antivortex binding energy, $\Phi_0$ is the flux quantum, $d$ is the interlayer spacing, $\lambda$ is the London penetration length depth, and $H_0 \sim H_{c2}$ (defined as $R_s(H_{c2}) = 0.9R_s(10\text{ K})$). These results indicate that the vortices in both mono- and bilayer devices exhibit thermally activated flux flow (TAFF). [29] From the fitting results of the bilayer device, we obtained $U_0/k_B = 9.1$ K and $H_0 = 0.37$ T ($H_0$ was found be smaller than $H_{c2}$ in this device), which are much larger than those of the monolayer device ($U_0/k_B = 0.12$ K and $H_0 = 0.37$ T). At lower temperatures, the resistance deviated from the thermally activated behavior, and a magnetic-field-induced metallic ground state emerged. For the bilayer device, we also measured the magnetic field dependence of $R_s$ at 0.3 K (see supplementary information for more details) and obtained a good fit (inset of Figure 3c) by using a model developed by Shimshoni et al [30] and Saito et al [17] in the low magnetic field regime, indicating that the vortices move through quantum tunneling.

The vortex phase diagrams of both mono- and bilayer devices based on our observations are plotted in Figure 3d. Two phases, the TAFF and quantum creep, are confirmed by the measured $H_{c2}$ and thermally activated behavior deviation points. When temperatures drop below $T_c$, the vortices move through the superconductor by thermal activation, whereas at even lower temperatures in a low magnetic field, vortices move by quantum tunneling. The movement of vortices indicates that even down to atomic thickness, the gate-induced 2D superconductor systems still reside in the weak disorder limit. In gate-induced superconductors in thick flakes of layered materials,[17] the superfluid density is confined in a few layers near the surface, and the 2D superconductors are susceptible only to relatively weak disorder generated by random electric potential from the ions; thus, they exhibit the behavior of a clean 2D superconductor. Herein, the mono- and bilayer $MoS_2$ flakes are exfoliated onto $SiO_2$ wafers, so the 2D superconductor is affected by stronger disorder from the substrate in addition to the ionic liquid. It is worth noting that these systems remain in a regime of a disordered 2D superconductor.

A quantum metallic state was also observed in bilayer $NbSe_2$ crystal covered by Boron nitride (BN).[11] The low temperature resistance of the state fulfills a power-law scaling with magnetic field, which is consistent with the so-called Bose-metal model. This mainly attributes to the covered BN which protects $NbSe_2$ crystal from the influence of atmosphere, resulting in a nearly disorder-free condition. While in the case of thin $MoS_2$ flakes, related EDLT systems remain in the disordered regime where quantum creep is expected to depict the quantum metallic state.[17] This explains why the low temperature resistance of these two systems exhibits different magnetic field dependence. For the quantum metallic state of $NbSe_2$ systems, it originates from the magnetic field induced strong phase fluctuations. Nevertheless, for the $MoS_2$ EDLT systems, magnetic-field-induced vortices move through quantum tunneling yields a different metallic ground state. More efforts should be involved to further understand the vortex dynamics of the $MoS_2$ EDLT systems.

**Discussion**

To fully understand the physical mechanism of 2D superconducting varying with carrier density, we perform *ab initio* density functional theory (DFT) calculations for the electronic structures of electron-doped monolayer and bilayer $MoS_2$. We choose several doping levels (from $0.57 \times 10^{14}$ cm$^{-2}$ to $2.74 \times 10^{14}$ cm$^{-2}$) to simulate the accumulated carrier density achieved in the liquid gating experiment, which is unavailable for conventional oxide-gated samples. It should be noted here that the electronic band structures of doped monolayer and bilayer $MoS_2$ strongly depend on the charge doping as shown in Figures 4a and 4d. Consistent with previous ARPES measurement, [31] the conduction band minima (CBM) of monolayer and bilayer $MoS_2$ are always located at the K (K') point even when electrons are injected into the samples. With increasing doping level, the conduction band edge at the K (K') point is filled, and the relative energy difference between the conduction band edge at the K (K') and Λ (Λ') points becomes smaller. By further increasing the electron doping, the states at the Λ (Λ') points are filled, whose energy splitting induced by the spin orbital coupling is much larger than that at the K (K') point. [32] Therefore, the electronic Fermi surface of the heavily doped $MoS_2$ contains two parts: one is around the K and K' points, and the other is located near the Λ and Λ' points. This result directly suggests that the electronic states around the Λ and Λ' points have a greater chance to be paired to form superconducting states via the phonon modes once the carrier density is sufficiently high to induce superconductivity in this system (as discussed below), which is different from previous reports. [20]

In the framework of Bardeen-Cooper-Schrieffer (BCS) theory, the superconducting $T_c$ is mainly determined by the electron-phonon coupling; [33] the averaged coupling constant has the expression of

$$\lambda = 2\int d\omega \frac{\alpha^2 F(\omega)}{\omega} = \sum_{q,v} \frac{2\gamma_{q,v}}{hN(\epsilon_F)\omega_{q,v}^2} \ , \tag{3}$$

where $\alpha^2 F(\omega)$ is the Eliashberg spectral functional, $N(\epsilon_F)$ is the electronic density of states (DOS) at the Fermi level, $\omega_{q,v}$ is the phonon frequency, and $\gamma_{q,v}$ is the phonon line width (see the Methods section). Thus, herein, by using density functional perturbation theory (DFPT), [34] we calculated the phonon bands and electron-phonon coupling for monolayer MoS$_2$ with different doping levels as shown in Figures 4b and 4c. Toward the low-frequency part of the acoustic phonons that mainly contributes to superconductivity, the phonon softening of longitudinal and transverse acoustic phonon modes (LA and TA modes) is found around the M point with low carrier density (blue line in Figure 4b); with increasing electron doping, the LA and TA modes can be further softened around the K point. This effect occurs because the electronic states around the Λ (Λ') point are filled at higher doping levels. However, for the whole calculated doping range, the out-of-plane mode (ZA mode), $N(\epsilon_F)$ and $\gamma_{q,v}$ change little (see supplementary information for more details). Thus, we conclude that the gate-induced superconductivity for the doped monolayer MoS$_2$ originates from the phonon softening of low-frequency TA and LA modes at the K and M points, and this result is further confirmed by the $\alpha^2F(\omega)$ calculations (see Figure 4c).

In contrast to monolayer MoS$_2$ superconductivity, the accumulated carrier density of a doped bilayer is lower for the same magnitude of $T_c$, which is observed experimentally in Fig. 2(c). These phenomena are confirmed by our DFPT calculations. In figures 4e and f, we show the calculated phonon dispersions and electron-phonon couplings for doped bilayer MoS$_2$. It is found that although the phonon softening also occurs at the M and K (K') points, the magnitude is much weaker in bilayer MoS$_2$ at the same doping level, which indicates that superconductivity in the bilayer or multilayer MoS$_2$ is more difficult to achieve due to the interlayer coupling. In experimental reality, the influence of the substrate and fluctuations in and inhomogeneity of the electron distribution also play important roles for 2D superconducting, which is still an open question and cannot be captured by our DFT calculations.

In conclusion, we demonstrate that in the gate-induced 2D superconductivity of both mono- and bilayer MoS$_2$ flakes, the monolayer sample has an apparently smaller critical sheet carrier density to achieve superconductivity than those of bilayer and bulk samples. The *ab initio* calculation results point to the phonon softening of in-plane acoustic phonon modes as a possible origin of these observations. Our work paves the way for further understanding of gated interfacial superconductivity approaching the ultimate atomic limit and pursuing a new type of superconductor.

**Methods**

Materials and devices

The mono- and bilayer $MoS_2$ flakes on silicon wafers (covered by 300-nm-thick $SiO_2$) were fabricated by standard mechanical exfoliation of bulk $MoS_2$ (SPI supplies). The number of layers was determined by measuring the thickness of the flakes using a Bruker Multimode 8 atomic force microscope (AFM) or performing micro Raman scattering (Horiba-JY T64000) measurements (under ambient conditions in the backscattering geometry with an incident laser wavelength of 514.5 nm). A conventional electron-beam lithography process (FEI F50 with an NPGS pattern generation system) followed by standard electron-beam evaporation of metal electrodes (typically 5 nm Ag/ 40 nm Au) was used to fabricate Hall bar electrodes.

Transport measurements

The ionic liquid used in this study was N;N-diethyl-N-(2-methoxyethyl) -N-methylammonium bis-(trifluoromethylsulfonyl)-imide (DEME-TFSI). An ionic liquid drop was applied onto $MoS_2$ devices with Hall geometry and covered both the $MoS_2$ flake and a side gate electrode to form an EDLT. Before any electrical measurements were performed, the $MoS_2$ EDLT devices were stored under a vacuum better than $10^{-2}$ mbar and cooled down to 220 K to avoid possible interfacial chemical reactions.

The transport measurements of all $MoS_2$ EDLTs were performed in an Oxford Instruments Teslatron$^{TM}$ CF cryostat. Due to the freezing of the ionic liquid at low temperatures, we applied different values of $V_{LG}$ to accumulate carriers of certain density at 220 K and cooled down with a fixed $V_{LG}$. A lock-in amplifier (Stanford Research 830) was used to measure the 4-probe resistance through the AC approach.

*Ab initio* calculations

The *ab initio* calculations were performed by using DFT as implemented in the Quantum Espresso [35] package. For the electronic structure calculations, the 32×32×1 Monkhorst-Pack *k* points were used with the generalized gradient approximation (GGA) functional [36] and norm-conserving pseudopotential. To correctly describe the electronic structures of monolayer and bilayer $MoS_2$ and repeat the previous calculations, [31] the in-plane lattice constant was chosen as 3.18 Å; the vacuum layer was chosen as 15 Å that was sufficiently large to avoid the interaction between adjacent layers, and inner coordinates of unit-cell were fully relaxed. Phonon band structures and electron-phonon couplings were calculated within density functional perturbation theory [34] based on the evaluation of the dynamical matrices on the 8×8×1 *q*-mesh. For the phonon dispersion, SOC was not considered.

**Acknowledgements**

This work was supported in part by the National Key Basic Research Program of China (2015CB921600, 2013CBA01603), the National Natural Science Foundation of China (11374142, 61574076), the Natural Science Foundation of Jiangsu Province (BK20130544, BK20140017, BK20150055), the Specialized Research Fund for the Doctoral Program of Higher Education (20130091120040), and Fundamental Research Funds for the Central Universities and the Collaborative Innovation Center of Advanced Microstructures. This work was partially supported by the Department of Energy, Office of Basic Energy Sciences, Division of Materials Sciences and Engineering, under contract DE-AC02-76SF00515. P.Z.T., B.L., G.X. and S.C.Z. also acknowledge NSF under grant number DMR-1305677.

## Data availability

The data that support the findings of this study are available from the corresponding author upon reasonable request.

## Competing financial interests

The authors declare no competing financial interests.

## Author contribution

Y.J. F., E.F. L., H.T. Y. and P.Z. T. contributed equally to this work. F. M. and H.T. Y. conceived the project and designed the experiments. Y.J. F., E.F. L., J.W. Z., Y.J. W., W. Z., K. X., A.Y. G., C. P., and M. W. performed device fabrication, characterization and transport measurements. Y.J. F., E.F. L., F. M. and H.T. Y. performed data analysis and interpretation. P.Z. T., B. L., X. G., Z.Y. C. and S.C. Z. performed DFT calculations. H.T. Y., F. M., Y.J. F. and E.F. L. co-wrote the paper, and all authors contributed to the discussion and preparation of the manuscript.

## Supplementary information

Supplementary information accompanies the paper on the *njp Quantum Materials* website.

28. Ephron, D., Yazdani, A., Kapitulnik, A. & Beasley, M. R. Observation of quantum dissipation in the vortex state of a highly disordered superconducting thin film. *Phys. Rev. Lett.* **76**, 1529-1532 (1996).

29. Feigel'Man, M. V., Geshkenbein, V. B. & Larkin, A. I. Pinning and creep in layered superconductors. *Physica C: Superconductivity* **167**, 177-187 (1990).

30. Shimshoni, E., Auerbach, A. & Kapitulnik, A. Transport through quantum melts. *Phys. Rev. Lett.* **80**, 3352-3355 (1998).

31. Yuan, H. T. *et al.* Evolution of the valley position in bulk transition-metal chalcogenides and their monolayer limit. *Nano Lett.* **16**, 4738-4745 (2016).

32. Molina-Sánchez, A., Sangalli, D., Hummer, K., Marini, A. & Wirtz, L. Effect of spin-orbit interaction on the optical spectra of single-layer, double-layer, and bulk $MoS_2$. *Phys. Rev. B* **88**, 045412 (2013).

33. Bardeen, J., Cooper, L. N. & Schrieffer, J. R. Theory of superconductivity. *Phys. Rev.* **108**, 1175-1204 (1957).

34. Baroni, S., De Gironcoli, S., Dal Corso, A. & Giannozzi, P. Phonons and related crystal properties from density-functional perturbation theory. *Rev. Mod. Phys.* **73**, 515-562 (2001).

35. Giannozzi, P. *et al.* QUANTUM ESPRESSO: a modular and open-source software project for quantum simulations of materials. *J. Phys.: Condens. Matter* **21**, 395502 (2009).

36. Perdew, J. P., Burke, K. & Ernzerhof, M. Generalized gradient approximation made simple. *Phys. Rev. Lett.* **77**, 3865-3868 (1996).
20

# Figure Captions

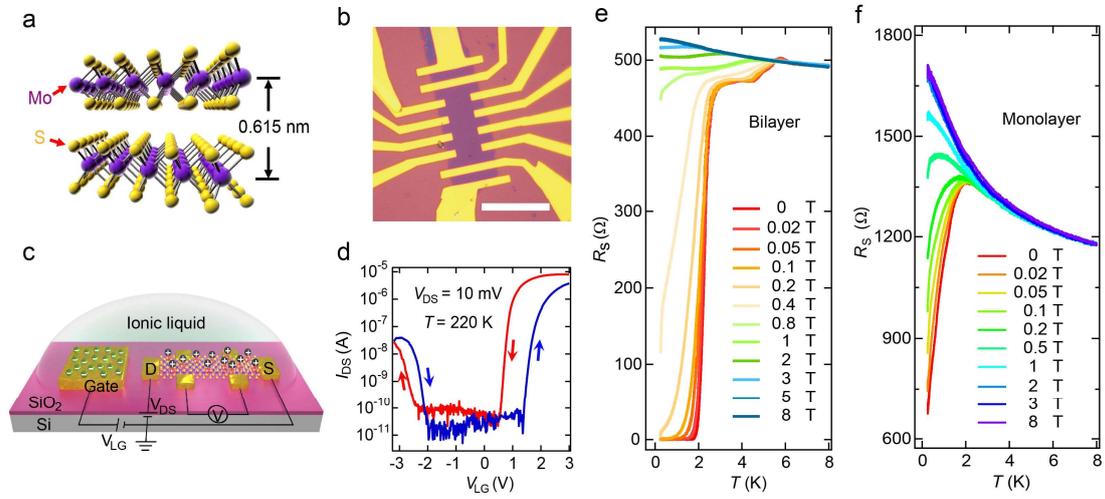

**Figure 1. Crystal structure, device geometry, and gate-induced superconductivity of mono- and bilayer MoS₂ EDLT devices. a**, Crystal structure of 2*H*-MoS$_2$, where the interlayer spacing is 0.615 nm. **b**, Optical microscope image of a typical bilayer MoS$_2$ EDLT device. Scale bar: 20 μm. **c**, Schematic illustration of an atomically thin MoS$_2$ EDLT device. The labels of S and D represent source and drain electrodes, respectively. **d**, Typical ambipolar transfer curves obtained from a bilayer MoS$_2$ EDLT device at 220 K (blue: forward direction; red: backward direction). $V_{DS}$ was fixed at 10 mV. **e, f**, *T*-dependent $R_s$ of bilayer (e) and monolayer (f) MoS$_2$ devices at different perpendicular magnetic fields showing gate-induced superconductivity.

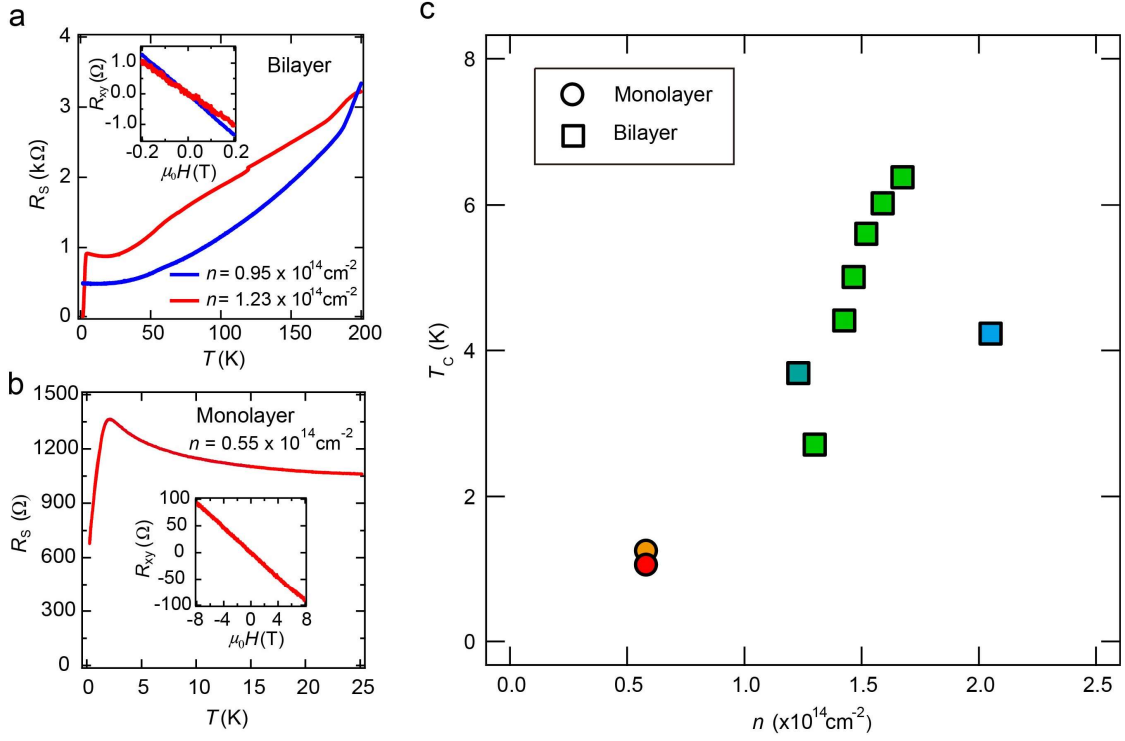

**Figure 2. Comparison of critical carrier densities to achieve superconductivity in mono- and bilayer MoS₂. a**, Temperature-dependent resistance of a bilayer MoS$_2$ EDLT at $n = 0.95 \times 10^{14}$ cm$^{-2}$ and $1.23 \times 10^{14}$ cm$^{-2}$. Inset: Hall measurement results. **b**, Temperature-dependent resistance of a monolayer MoS$_2$ EDLT at $n = 0.55 \times 10^{14}$ cm$^{-2}$. Inset: Hall measurement results. **c**, Critical temperature $T_c$ versus carrier density $n$ for mono- and bilayer MoS$_2$. $T_c$ is defined as $R_s(T_c) = 0.9 R_s(10\ \mathrm{K})$.

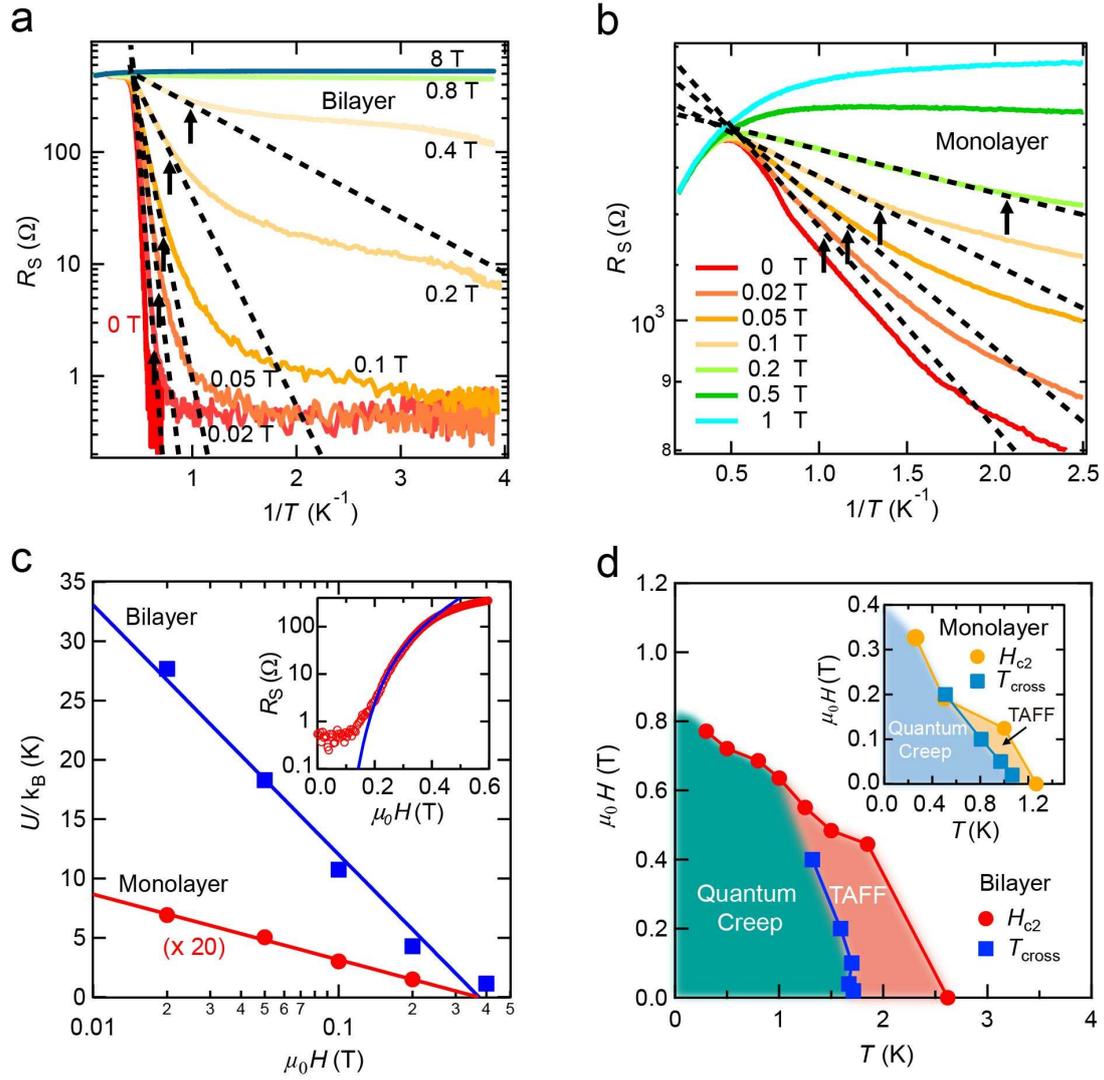

**Figure 3. Vortex dynamics analysis of mono- and bilayer MoS$_2$. a** and **b** show the Arrhenius plot of the sheet resistance of two typical bi- and monolayer MoS$_2$ EDLTs at different perpendicular magnetic fields, respectively. The dashed lines show the guide-to-eye fit of the thermally activated behavior, and the arrows show the deviation of the thermally activated regimes. **c**, Semilogarithmic plot of the activation energy, $U(H)/k_B$, as a function of magnetic field for both mono- and bilayer devices. $U(H)/k_B$ data were extracted from fittings in Figures 3a and 3b. The solid line is a fit to the formula $U(H)=U_0\ln(H_0/H)$. Inset: the magnetic field dependence of $R_s$ at 0.3 K with a fit obtained by using a model developed by Shimshoni *et al* [30] and Saito *et al*. [17] **d**, Vortex phase diagram of the bi- and monolayer (inset) MoS$_2$ EDLT devices. The solid circles represent $H_{c2}$ at different temperatures, and the solid squares were extracted from the deviation points as shown in **a** and **b** (depicted by the black arrows). The solid squares divide the thermally activated flux flow (TAFF) and quantum creep regimes.

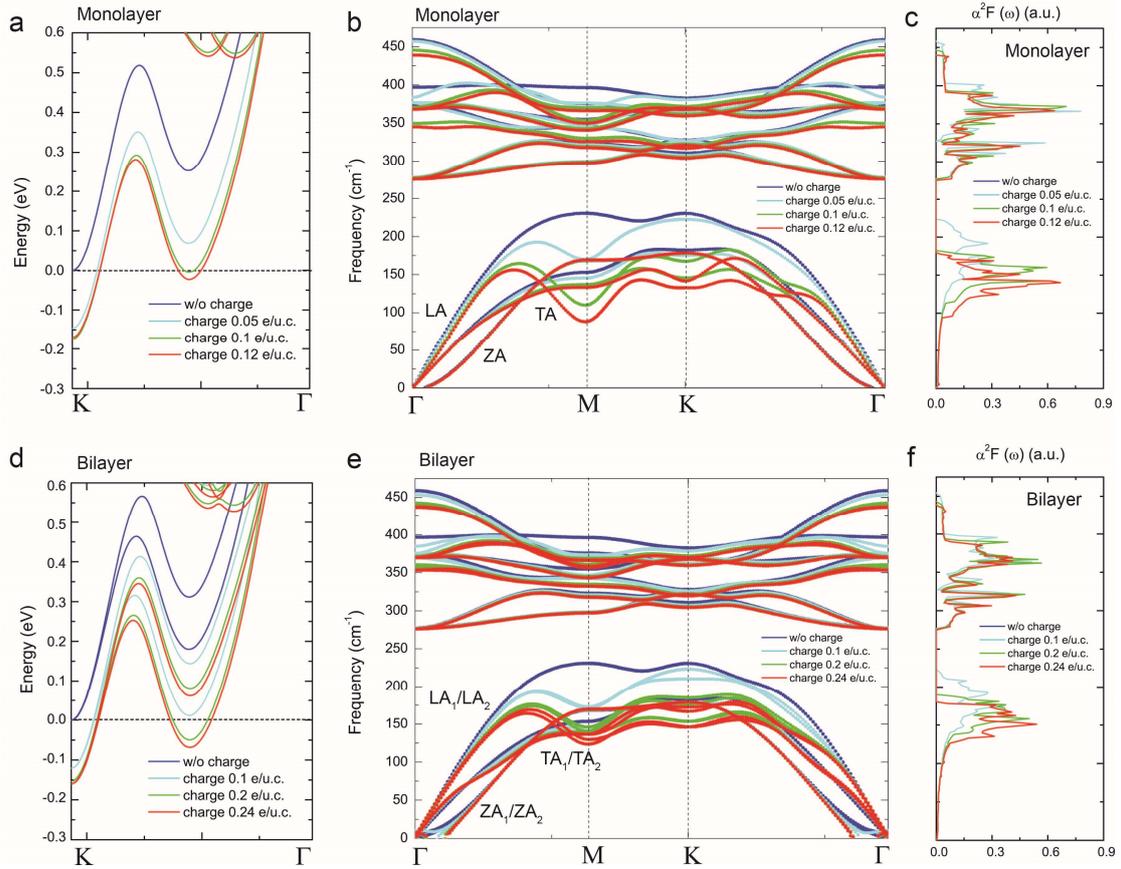

**Figure 4. Calculated electronic structures and electron-phonon couplings for the doped mono- and bilayer MoS$_2$.** The band structures of **a** monolayer and **d** bilayer MoS$_2$ with different doping levels. The Fermi level is set to zero and marked by black dotted lines. For the un-doped case, we set the energy of the conduction band minimum to zero. The phonon dispersion and the Eliashberg spectral functional for **b, c** monolayer and **e, f** bilayer MoS$_2$ with different doping levels. Herein, LA, TA and ZA stand for in-plane longitudinal mode, in-plane transverse mode and out-plane transverse mode. For bilayer MoS$_2$, the polarizations of these pairs of acoustic branches are denoted as follows: LA$_1$/LA$_2$, TA$_1$/TA$_2$ and ZA$_1$/ZA$_2$.

# Supplementary Information for

# Gated Tuned Superconductivity and Phonon Softening in Mono- and Bilayer MoS$_2$

Yajun Fu[1]*, Erfu Liu[1]*, Hongtao Yuan[2,3]*, Peizhe Tang[3]*, Biao Lian[3], Gang Xu[3], Junwen Zeng[1], Zhuoyu Chen[3], Yaojia Wang[1], Wei Zhou[1], Kang Xu[1], Anyuan Gao[1], Chen Pan[1], Miao Wang[1], Baigeng Wang[1], Shou-Cheng Zhang[2,3], Yi Cui[2,3], Harold Y. Hwang[2,3] & Feng Miao[1]

[1]National Laboratory of Solid State Microstructures, School of Physics, Collaborative Innovation Center of Advanced Microstructures, Nanjing University, Nanjing 210093, China.

[2]Stanford Institute for Materials and Energy Sciences, SLAC National Accelerator Laboratory, Menlo Park, California 94025, USA.

[3]Geballe Laboratory for Advanced Materials, Stanford University, Stanford, California 94305, USA.

**Supplementary Note 1: Identification of layer number**

The thickness of the as-exfoliated mono- and bilayer MoS$_2$ flakes were identified by micro Raman spectroscopy. The separation of $E_{2g}^1$ and $A_{1g}$ peaks was measured to be 18.4 and 21.4 cm$^{-1}$ for mono- and bilayer MoS$_2$, respectively (Fig. S1a), consistent with the previous reports.[1,2] After the low-temperature ionic-gating transport measurements were performed, the post-measured devices showed Raman characteristics identical to those of the previously measured flakes, indicating the absence of a phase transition (2H-1T), which might occur with the ultrahigh carrier density doping. As shown in Fig. S1b and S1c, no additional peaks were observed in the post-measured devices at approximately 156 cm$^{-1}$, 226 cm$^{-1}$ and 330 cm$^{-1}$ (named J1, J2 and J3, respectively, for 1T-MoS$_2$).[3,4]

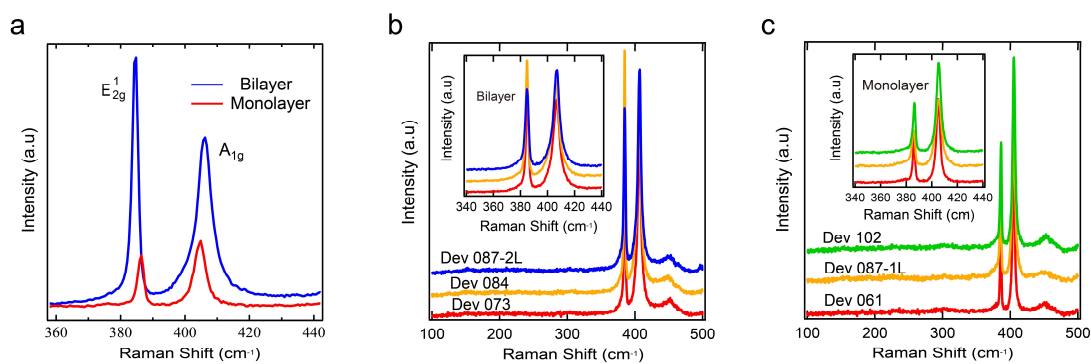

**Figure S1 | Raman spectra of mono- and bilayer MoS$_2$. a**, Raman spectra of as-exfoliated mono- and bilayer MoS$_2$ flakes. **b**, and **c**, Raman spectra of the post-measured bi- and monolayer MoS$_2$ devices, indicating the absence of the 2H-1T phase transition. Inset: zoomed-in data of $E_{2g}^1$ and $A_{1g}$ peaks.

**Supplementary Note 2: $R_s$-$T$ curves and magnetoresistance data**

Fig. S2a shows an $R_s$-$T$ curve of a bilayer $MoS_2$ device with a larger temperature range (the same device as shown in Fig. 1e) with $n = 1.3 \times 10^{14}$ cm$^{-2}$ (measured at $T = 10$ K). A metal-insulator transition was observed at ~15 K, and a superconducting state was observed when $T < 1.7$ K. As shown in Fig. S2b, by applying a perpendicular magnetic field at a base temperature (300 mK), the zero resistance state could be suppressed starting with $\mu_0 H \sim 0.15$ T and recovered to the normal state with $\mu_0 H = 1$ T. Fig. S2c shows the magnetoresistance measurements of a monolayer device (the same device as shown in Fig. 1f) at different temperatures.

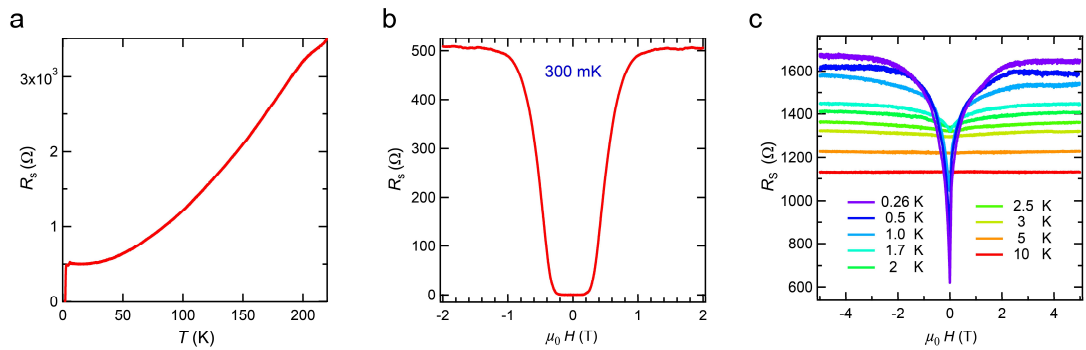

**Figure S2 | $R_s$-$T$ curves and magnetoresistance data. a**, Larger temperature range $R_s$-$T$ curve of a bilayer $MoS_2$ device (the same device as shown in Fig. 1e). **b**, Magnetoresistance measurement performed at 0.3 K. **c**, Magnetoresistance measurements of a monolayer device (the same device as shown in Fig. 1f) at different temperatures.

**Supplementary Note 3: *T*- and *B*-dependent *V-I* characteristics of bilayer MoS$_2$**

We measured the 4-probe voltage-current (*V-I*) characteristics using a standard four-terminal dc technique. As shown by the results obtained from a typical bilayer MoS$_2$ device (Fig. S3a and S3b), a well-defined critical current $I_c$ ~11.09 µA was observed at a base temperature of 280 mK. The zero resistance state was suppressed gradually with either increased temperature (Fig. S3a) or the application of perpendicular magnetic field *B* (Fig. S3b). A complete suppression (showing linear *V-I* curves) was observed when $T \geq 3$ K or $\mu_0 H \geq 8$ T. Fig. S3c shows the color plot of *V-I* curves with continual sweeping of *B* (ranging from -1 T to 1 T).

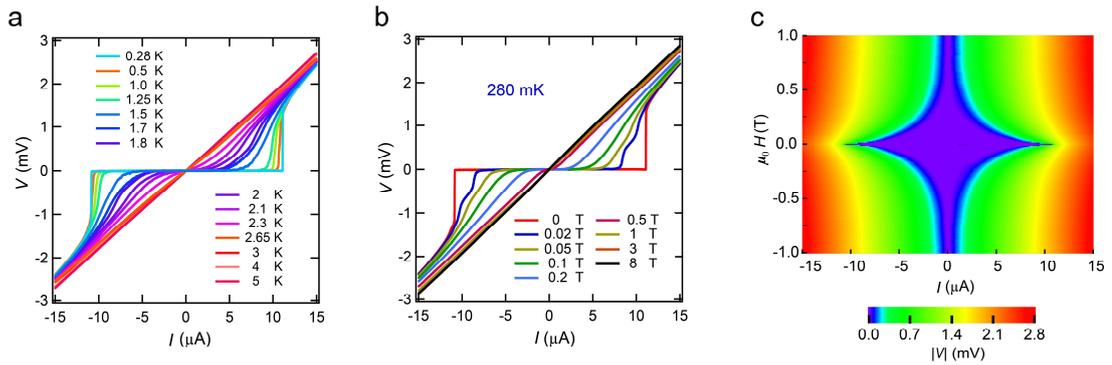

**Figure S3 | Voltage-current (*V-I*) characteristics of bilayer MoS$_2$. a**, *V-I* curves at different temperatures (0.28 – 5 K) with zero magnetic field. **b**, *V-I* curves at different magnetic fields (0 – 8 T) at 280 mK. **c**, Color plot of *V-I* curves with continual sweeping of *B* (ranging from -1 T to 1 T). The different colors represent the magnitude of the voltage.

**Supplementary Note 4: BKT transition of bilayer MoS$_2$**

As shown in Fig. S4a, a clear $V \sim I^\alpha$ power law dependence was observed in a bilayer MoS$_2$ device ($n = 1.3 \times 10^{14}$ cm$^{-2}$) at various temperatures ranging from 0.28 to 5 K, suggesting the occurrence of the Berezinskii-Kosterlitz-Thouless (BKT) transition [5-7] for 2D superconductivity. As shown in Fig. S4b, the fitted exponent $\alpha$ decreased with increasing temperature and approached 3 at $T \sim 2$ K, indicating a BKT transition temperature $T_{BKT} \sim 2$ K. An additional test for the BKT transition is to check whether $[d(\ln R)/dT]^{-2/3}$ varies linearly with $T$ when above $T_{BKT}$, [5-7] which was also observed (shown in Fig. S4c). $T_{BKT}$ was determined to be ~2.17 K from extrapolation, consistent with the previous estimation (~2 K) derived in Fig. S4b.

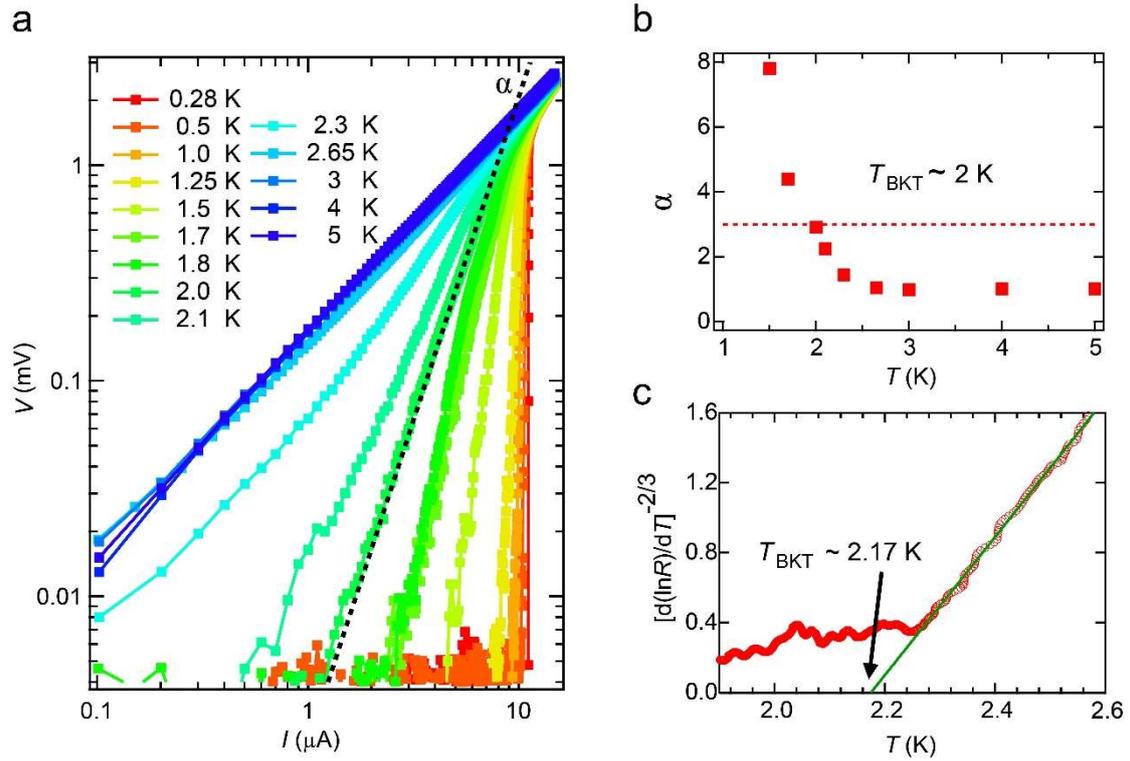

**Figure S4 | BKT transition of bilayer MoS$_2$ superconductivity. a**, $T$-dependent $V$-$I$ curves plotted on a logarithmic scale. The dashed black line corresponds to $V \propto I^3$. **b**, The power law exponent $\alpha$ (extracted from Fig. S4a) versus $T$. The dashed line corresponds to $\alpha = 3$ and determines that $T_{BKT} \sim 2$ K. **c**, $[d(\ln R)/dT]^{-2/3}$ versus $T$. The green line shows the extrapolation of the linear part and defines $T_{BKT}$ as ~2.17 K.

**Supplementary Note 5: Enhanced back gate modulation of superconductivity of bilayer MoS$_2$**

Additional experimental evidence for the 2D nature of the ion-liquid gating-induced superconductivity in bilayer MoS$_2$ is the observation of the enhanced back gate (with 300 nm SiO$_2$ as a dielectric) modulation. Fig. S5a shows the results of $T$-dependent $R_s$ measurements at different back gate voltages ($V_{bg}$) with $n = 1.52 \times 10^{14}$ cm$^{-2}$ (measured at 10 K). In addition to the normal state $R_s$, the superconducting critical temperature $T_c$ (defined as $R_s(T_c) = 0.9R_s(10\text{ K})$) was also found to be notably modulated by $V_{bg}$. As shown in Fig. S5b, whereas $V_{bg}$ varied from -40 V to 40 V, $T_c$ increased from ~4.4 to ~6.4 K, and the measured carrier density $n$ was modulated from $1.43 \times 10^{14}$ cm$^{-2}$ to $1.67 \times 10^{14}$ cm$^{-2}$, which is a significantly enhanced modulation compared with thicker MoS$_2$ devices (~20 nm). [8] This effect is mainly due to the modulatory effects of both ion-liquid and back gates on the $n$ of a given 2D electron gas (2DEG) upon approaching the ultimate 2D limit. By contrast, in the case of thicker samples, two 2DEGs are separately formed on the top and bottom surfaces. More interestingly, in such an ultrathin MoS$_2$ channel where the electron accumulation is fully confined inside the atomically thin flakes, the supercurrent can be switched on and off by tuning the back gate voltage only (without tuning the liquid gate voltage), as shown in Fig. S5c. $H_{c2}$ (defined as $R_s(H_{c2}) = 0.9R_s(10\text{ K})$) can also be modulated significantly, as shown in Fig. S5d. Our results suggest that in devices approaching the 2D limit, back gating with a dual gate geometry could be very powerful for modulating superconductivity.

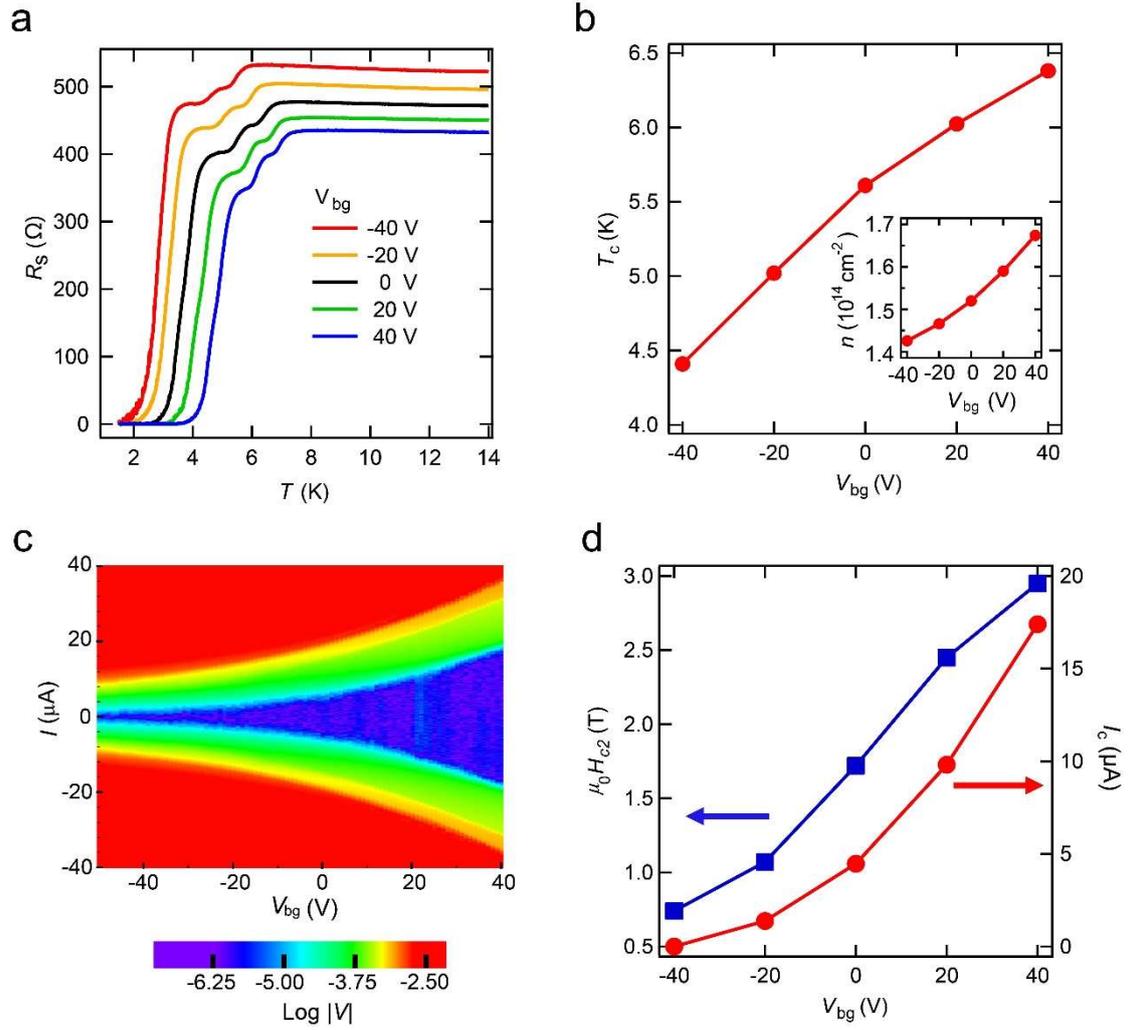

**Figure S5 | Back gate voltage-tunable superconducting transition. a**, $T$-dependent $R_s$ for different back gate voltages $V_{bg}$. The carrier density was measured to be $1.52 \times 10^{14}$ cm$^{-2}$ when $V_{bg} = 0$ V. **b**, $V_{bg}$ dependent critical temperature $T_c$, where $T_c$ is defined as $R_s(T_c) = 0.9R_s(10\ \text{K})$. Inset: carrier density versus back gate voltage measured at 10 K. **c**, Color plot (in logarithmic scale) of $V$-$I$-$V_{bg}$, indicating realization of the switching of the supercurrent tuned by $V_{bg}$ only. **d**, The critical current $I_c$ and critical magnetic field $\mu_0 H_{c2}$ as a function of $V_{bg}$ at 2 K.

**Supplementary Note 6: Plot using number of carrier per primitive cell as the standard of the critical carrier density**

In MoS$_2$ EDLT devices, $n_{2D}$ of individual layer decays exponentially from the channel surface, reducing $n_{2D}$ of the second-to-outermost layer by almost 90% in comparison with the outermost one. [8, 9] In our bilayer MoS2 case, the 2D carrier density showed in the main text were ~$10^{14}$ cm$^{-2}$ in total, and we consider that most of the carriers were accumulated in the top monolayer as well. Therefore, it is reasonable to use the sheet carrier density as the standard of the critical carrier density for inducing superconductivity.

The number of carriers per primitive cell can be used as another meaningful standard of the critical carrier density. Such plot is shown in Fig. S6. Same conclusion can be reached that the critical carrier density for inducing superconductivity in monolayer MoS$_2$ (~0.05 electron per primitive cell) is much lower than those in the bilayer case (~0.1 electron per primitive cell).

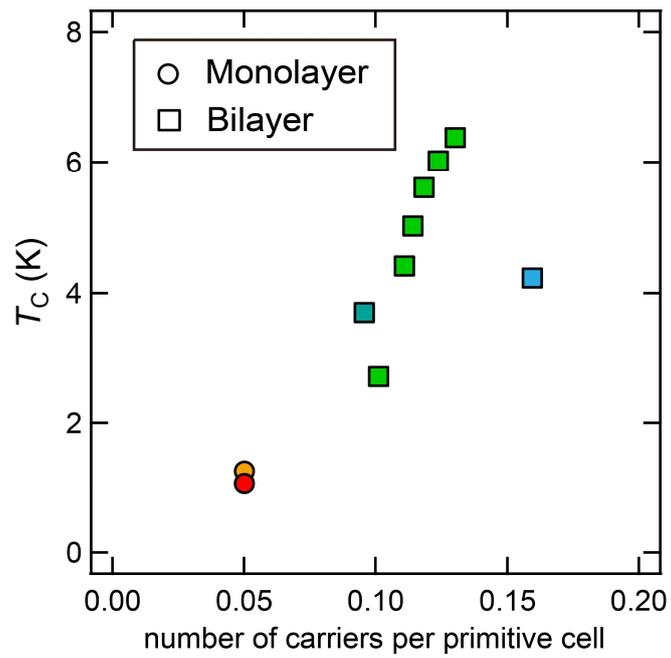

**Figure S6 | Plot using number of carrier per primitive cell as the standard of the critical carrier density**. Critical temperature $T_c$ versus number of carrier per primitive cell for mono- and bilayer $MoS_2$.

**Supplementary Note 7: Homogeneity of carrier densities for multiple pairs of contacts**

Fig. S7a shows the picture of a typical monolayer flake with multi-probe configuration yielding multiple pairs of contacts. Most contacts survived from the cooling processes except contact #7. We observed superconducting behaviors for the pairs 2-3, 3-4 and 8-9 simultaneously, with results shown in Fig. S7b. We further performed dc Hall effect measurements of pairs 2-9 and 3-8 at 10 K (Fig. S7c), and measured the corresponding carrier densities to be $5.5\times10^{13}$ cm$^{-2}$ and $5.8\times10^{13}$ cm$^{-2}$, respectively.

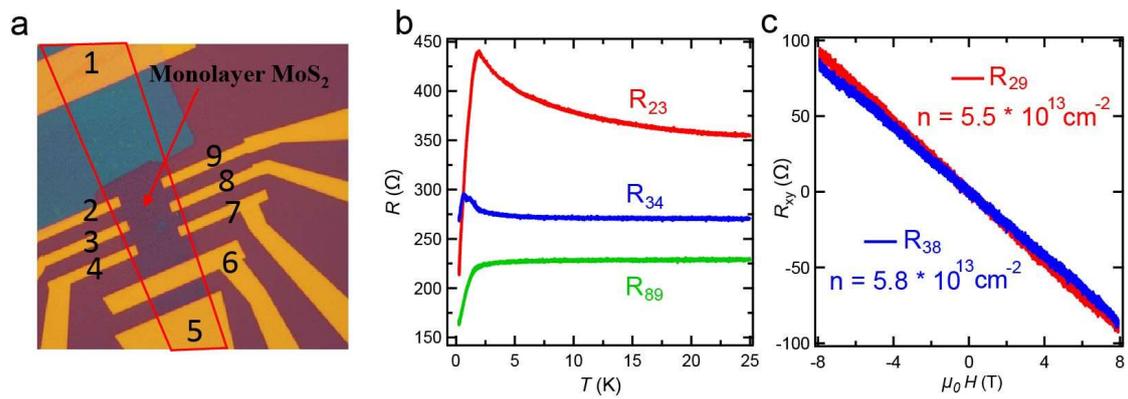

**Figure S7 | Device geometry, gate-induced superconductivity and Hall effect of a typical monolayer MoS$_2$ flake.** a, Optical microscope image of the flake with multi-probe configuration (different electrodes labelled by numbers); b, $T$-dependent resistance of different electrode pairs; c, Hall effect measurement of electrode pairs 2-9 and 3-8 at 10 K. The calculated carrier densities for each pair are marked.

**Supplementary Note 8: Gating-induced Raman peak shift of monolayer MoS$_2$**

Fig. S8a shows the evolution of the E$^1_{2g}$ and A$_{1g}$ modes of a MoS$_2$ monolayer at different ionic gate voltages $V_{LG}$ (measured at 290 K). Fig. S8b plots the shift in the A$_{1g}$ peak versus $V_{LG}$. When $V_{LG}$ approached 3 V, the A$_{1g}$ mode frequency was observed to be softened by up to ~ 7 cm$^{-1}$. On the other hand, the E$^1_{2g}$ mode also slightly red shifted under high gate voltages (Fig. S8c). The observation of the softening of the Raman spectra induced by $V_{LG}$ (i.e., electron accumulation) suggests a strong electron-phonon coupling and will stimulates the understanding of the superconducting mechanism of the ionic liquid-gated monolayer MoS$_2$.

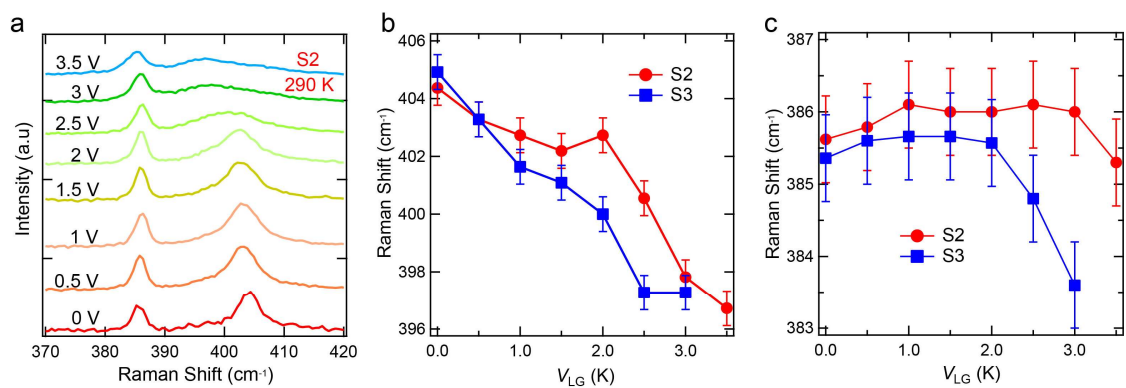

**Figure S8 | Ionic gating-induced Raman peak shift of monolayer MoS$_2$. a**, The Raman spectra of monolayer MoS$_2$ at different ionic gate voltages $V_{LG}$ (measured at 290 K). **b**, Shift in the A$_{1g}$ peak for two different samples at different $V_{LG}$. Error bars were determined by the measurement system. **c**, Shift in the E$^1_{2g}$ peak for two different samples at different $V_{LG}$. Error bars were determined by the measurement system.

**Supplementary Note 9: Superconducting temperature from BCS theory**

Herein, the superconducting properties were calculated based on the Eliashberg theory, [10] in which the Eliashberg spectral functional could be expressed as

$$\alpha^2 F(\omega) = \frac{1}{2\pi N(\epsilon_F)} \sum_{q,v} \delta(\omega - \omega_{q,v}) \frac{2\pi \gamma_{q,v}}{\hbar \omega_{q,v}}$$

where $N(\epsilon_F)$ is the electronic DOS at the Fermi level, $\omega_{q,v}$ is the phonon frequency, and $\gamma_{q,v}$ is the phonon linewidth. It can be expressed as

$$\gamma_{q,v} = 2\pi \omega_{q,v} \sum_{ij} \int \frac{d^3k}{\Omega_{BZ}} |g_{q,v}(k,i,j)|^2 \delta(\epsilon_{k,i} - \epsilon_F)\delta(\epsilon_{k+q,j} - \epsilon_F)$$

where $g_{q,v}(k,i,j)$ is the electron-phonon coupling matrix elements, $\epsilon_F$ is the Fermi surface energy, and $\epsilon_{k,i}$ and $\epsilon_{k+q,j}$ are the energies of the electronic states to be paired. $\Omega_{BZ}$ is the volume of the Brillouin zone (BZ).

The superconducting temperature $T_c$ can be evaluated via the Allen-Dynes formula:

$$T_c = \frac{\hbar \omega_{log}}{2\pi \times 1.2} \exp[\frac{-1.04(1+\lambda)}{\lambda(1 - 0.62\mu^*) - \mu^*}]$$

where

$$\lambda = 2 \int d\omega \frac{\alpha^2 F(\omega)}{\omega} = \sum_{q,v} \frac{2\gamma_{q,v}}{\hbar N(\epsilon_F)\omega_{q,v}^2}$$

is the averaged electron-phonon coupling constant,

$$\omega_{log} = \exp[\frac{2}{\lambda} \int d\omega \alpha^2 F(\omega) \frac{\log(\omega)}{\omega}]$$

is the logarithmic averaged typical characteristic phonon frequency (K), and $\mu^*$ is the effective Coulomb potential.

**Supplementary Note 10: The phonon linewidth for monolayer and bilayer MoS$_2$ with charge doping**

Fig. S9a shows the calculated phonon linewidth of monolayer MoS$_2$ with charge doping from 0.05 e/u.c. to 0.12 e/u.c. The phonon linewidth did not change very much with increasing charge doping, especially for the ZA mode. As explained above, the electron-phonon coupling constant $\lambda_{q,\nu}$ of mode ν is proportional to $\frac{2\gamma_{q,\nu}}{hN(\epsilon_F)\omega_{q,\nu}^2}$, and $\lambda_{q,\nu}$ is strongly enhanced when charge carriers are injected into MoS$_2$ samples. If the phonon linewidth is independent of the charge doping, the enhancement of superconductivity for charged monolayer MoS$_2$ is mainly contributed by the phonon softening because the change in DOS at the Fermi level is not dramatic. Fig. S9b shows a plot of the phonon linewidth of bilayer MoS$_2$ with charge doping from 0.1 e/u.c. to 0.24 e/u.c. Similar to the monolayer case, the phonon linewidths of acoustic phonon modes in bilayer MoS$_2$ do not vary significantly with electron doping. Moreover, their magnitudes are smaller than those in monolayer MoS$_2$, which indicates that electron-phonon coupling in bilayer MoS$_2$ is weaker than that in doped monolayer MoS$_2$.

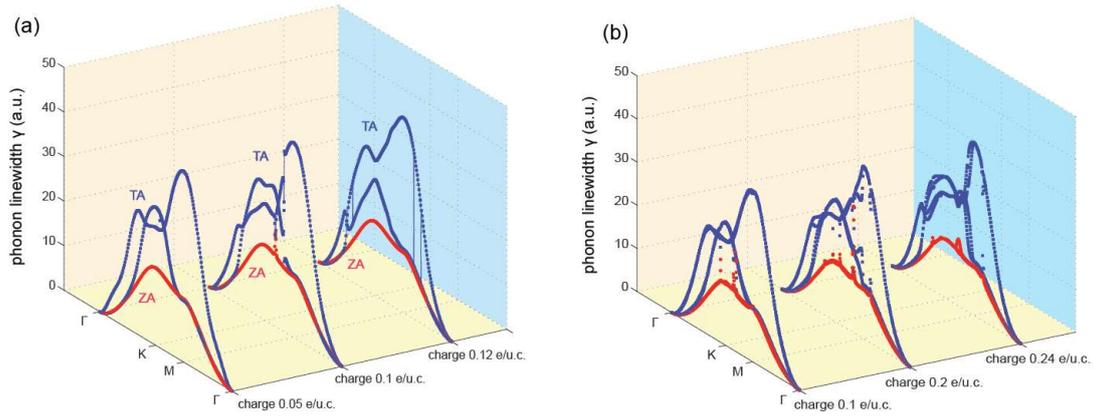

**Figure S9 | Phonon linewidth of monolayer and bilayer MoS₂ with charge doping. a**, The phonon linewidth of monolayer MoS$_2$ with charge doping from 0.05 e/u.c. to 0.12 e/u.c along the high symmetric lines in the BZ. The red and blue lines represent the contribution from the ZA mode and LA mode, respectively. **b**, The phonon linewidth of bilayer MoS$_2$ with charge doping from 0.1 e/u.c. to 0.24 e/u.c along the high symmetric lines in the BZ. The red lines are contributed by the lowest two branches of modes, and the blue lines are contributed by the other acoustic phonon modes.